\documentclass[%
 aps,
 amsmath,amssymb,reprint,
 pra,superscriptaddress]{revtex4-2}
\usepackage{mathrsfs}
\usepackage{graphicx}
\usepackage{dcolumn}
\usepackage{bm}
\usepackage{pifont} 
\usepackage{siunitx}
\usepackage[normalem]{ulem}

\DeclareSIUnit\gauss{G}
\DeclareSIUnit{\au}{a.u.}

\usepackage[usenames,dvipsnames]{xcolor}
\usepackage[colorlinks=true,urlcolor=blue,citecolor=magenta,linkcolor=magenta]{hyperref}
\usepackage[final]{microtype}
\usepackage[dvipsnames]{xcolor}

\usepackage{orcidlink}
\usepackage{nicematrix}
\NiceMatrixOptions{
code-for-first-row = \color{black!30!ForestGreen} ,
code-for-last-row = \color{black!30!ForestGreen} ,
code-for-first-col = \color{black!30!ForestGreen} ,
code-for-last-col = \color{black!30!ForestGreen}
}

\makeatletter
\newcommand{\ORCID}[1]{{\orcidlink{#1}}}
\makeatother

\newcommand{\Rb}{$\rm^{87}Rb$}

\newcommand{\ket}[1]{\ensuremath{\left|#1\right>}}
\newcommand{\bra}[1]{\ensuremath{\left<#1\right|}}
\newcommand{\ReducedMat}[3]{\ensuremath{\left<\left. #1 \right\| #2 \left\| #3 \right.\right>}}

\newcommand{\eref}[1]{Eq.~(\ref{#1})}
\newcommand{\Eref}[1]{Equation (\ref{#1})}

\newcommand{\fref}[1]{Fig.~\ref{#1}}
\newcommand{\figref}[1]{Fig.~\ref{#1}}
\newcommand{\Fref}[1]{Figure~\ref{#1}}

\newcommand{\thetaMW}{\ensuremath{\theta_{\text{MW}}}}

\begin{document}
\preprint{AIP/123-QED}

\title{Polarization-insensitive microwave electrometry using Rydberg atoms}
\author{Matthew Cloutman}
\affiliation{
Department of Physics, QSO—Quantum Science Otago, and Dodd-Walls Centre for Photonic and Quantum Technologies,
University of Otago, Dunedin 9016, New Zealand
}%
\author{Matthew Chilcott\ORCID{0000-0002-1664-6477}}
\affiliation{
Department of Physics, QSO—Quantum Science Otago, and Dodd-Walls Centre for Photonic and Quantum Technologies,
University of Otago, Dunedin 9016, New Zealand
}%
\author{Alexander Elliott}
\affiliation{
Department of Physics and Dodd-Walls Centre for Photonic and Quantum Technologies,
University of Auckland, New Zealand
}%
\author{J. Susanne Otto\ORCID{0000-0003-0760-3800}}
\affiliation{
Department of Physics, QSO—Quantum Science Otago, and Dodd-Walls Centre for Photonic and Quantum Technologies,
University of Otago, Dunedin 9016, New Zealand
}%
\author{Amita B. Deb\ORCID{0000-0002-2427-3500}}%
\affiliation{
School of Physics and Astronomy, University of Birmingham, Edgbaston, Birmingham B15 2TT, United Kingdom
}%
\author{Niels Kj{\ae}rgaard\ORCID{0000-0002-7830-9468}}%
 \email{niels.kjaergaard@otago.ac.nz}
\affiliation{
Department of Physics, QSO—Quantum Science Otago, and Dodd-Walls Centre for Photonic and Quantum Technologies,
University of Otago, Dunedin 9016, New Zealand
}%
\date{\today}

\begin{abstract}
We investigate the Autler-Townes splitting for Rydberg atoms dressed with linearly polarized  microwave radiation, resonant with generic $S_{1/2}\leftrightarrow{P}_{1/2}$ and $S_{1/2}\leftrightarrow{P}_{3/2}$ transitions. The splitting is probed using laser light via electromagnetically-induced transparency measurements, where the transmission of probe laser light reveals a two-peak pattern. In particular, this pattern is invariant under rotation of the microwave field polarization. In consequence, we establish $S \leftrightarrow P$ Rydberg transitions as ideally suited for polarization-insensitive electrometry, contrary to recent findings [A.~Chopinaud and J.~D.~Pritchard, \href{https://journals.aps.org/prapplied/abstract/10.1103/PhysRevApplied.16.024008}{Phys. Rev. Appl. {\bf 16}, 024008 (2021)}].
\end{abstract}

\maketitle
\section{Introduction}
\begin{figure*}[bt!]
    \centering
    \includegraphics[width=\linewidth]{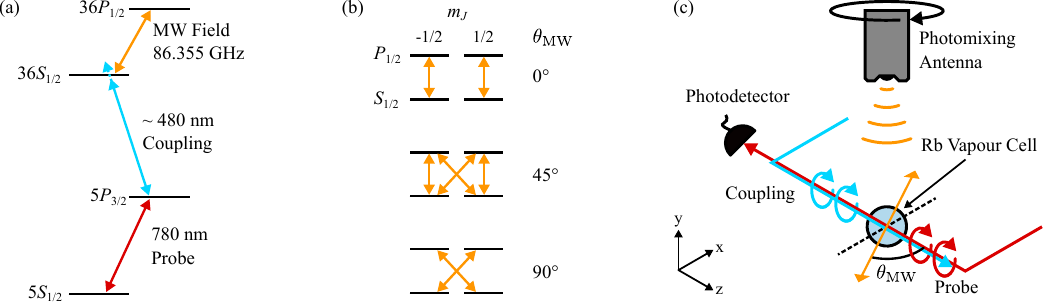}
    \caption{(a) Reduced $\rm ^{87}Rb$ level diagram. An infrared probe laser is locked to the $5S_{1/2} \leftrightarrow 5P_{3/2}$ D2 line while a blue coupling laser scans across the $5P_{3/2} \leftrightarrow 36S_{1/2}$ transition. A microwave field couples the $36S_{1/2}$ and $36P_{1/2}$ levels separated by \SI{86.355}{GHz}. (b) The $m_J$ magnetic sublevels involved in the $36S_{1/2} \leftrightarrow 36P_{1/2}$ transition and their coupling for \SI{0}{\degree}, \SI{45}{\degree}, and \SI{90}{\degree} polarization angles. (c) Experimental setup. Probe and coupling laser beams are circularly polarized and counter-propagate through a \SI{10}{mm} diameter bubble cell filled with a rubidium vapour. The emitter is mounted in a motorised rotation stage and emits a linearly polarized field.}
    \label{fig:pol-spec-diagram}
\end{figure*}
Rydberg atom-based electric field sensors offer high sensitivities \cite{Jing_2020,Prajapati2021}, large frequency bandwidth \cite{Hu2022}, SI-traceability, and self-calibration \cite{Sedlacek2012,Holloway2014,Anderson21}. Highly excited Rydberg atoms have large transition dipole moments to nearby Rydberg states, which can be resonantly addressed with radiation in the microwave (MW) and terahertz domains. This makes Rydberg atoms responsive to a wide frequency range and their high sensitivity to electric fields is being harnessed for a growing number of applications, as for example in precision metrology \cite{Holloway2017,Holloway2017b}, communication \cite{Meyer2018,Song2018,Deb2018,Otto2023}, imaging \cite{Holloway2017c,Downes2020,Anderson2023}, and measurements of phase \cite{Anderson2022,berweger2023} and polarization \cite{Sedlacek2013,Jiao2017,Anderson2018,Wang23}.

To probe the effect of an electric field on the Rydberg atomic medium, the atoms are typically optically interrogated by two laser fields using electromagnetically-induced transparency (EIT) \cite{Mohapatra2007}. In particular, if the applied electric field is resonant with the transition frequency between two Rydberg states, a splitting of the EIT signal emerges known as Autler-Townes (AT) splitting. The electric field strength can be directly retrieved from the AT splitting of the EIT signal \cite{SIMONS2021100273}.  
In a recent study, Chopinaud and Pritchard considered optimal atomic state choices for robust microwave measurements~\cite{Chopinaud2021}. One conclusion drawn in Ref.~\onlinecite{Chopinaud2021} is that the AT splitting resulting from linearly polarized microwaves driving transitions between $S_{1/2}$ and $P_{1/2}$ Rydberg levels is highly sensitive to the angle of the microwave field polarization. Furthermore, the study, which is underpinned by experiments, claims to find a ``complex AT splitting that can be used to extract a polarization ratio using an alternative method to that of [Ref.~\cite{Sedlacek2013}]'', and predicts crossings and shifting of the AT peaks.

In this paper, we show that an atom interacting with a linearly polarized microwave field gives rise to (dressed) eigenenergies with values that remain constant upon rotating the polarization. That the eigenenergies  cannot change can be inferred from a symmetry argument where the quantization axis is chosen along the direction of polarization. Because of the rotational symmetry of the atomic system, no matter the orientation of the field polarization, the atom-field coupling will always be described by the same series of $\pi$-transitions. For the $S_{1/2} \leftrightarrow P_{1/2}$ transition resonantly dressed with a linearly polarized microwave field, Ref.~\cite{Chopinaud2021} finds up to four eigenenergies, but as we show below there are only two distinct eigenenergies, each two-fold degenerate. We additionally provide an experimental demonstration that a $S_{1/2} \leftrightarrow P_{1/2}$ transition dressed by a microwave field indeed leads to two AT peaks when probed in an EIT ladder-scheme. Moreover, the AT peak positions are insensitive to the rotation of the dressing field polarization.

\section{Polarization-independent eigenenergies}
The experiments of Ref.~\cite{Chopinaud2021} considered  cesium atoms dressed in linearly polarized microwave radiation, resonant with the transition between the two Rydberg levels $65S_{1/2}$ and $65{P}_{1/2}$. Our experiments presented in this paper make use of the $36S_{1/2}$ and $36{P}_{1/2}$ levels of \Rb, which establishes an equivalent scenario as it is the coupled $J=1/2\Leftrightarrow J'=1/2$ angular momentum characteristics (factored out by the Wigner-Eckardt theorem) which contains the relevant physics. \Fref{fig:pol-spec-diagram}(a) shows a level diagram for $\rm ^{87}Rb$  while \fref{fig:pol-spec-diagram}(b) shows how the $S_{1/2}$ and ${P}_{1/2}$ magnetic sublevels are coupled for linearly polarized MW at angles $0^\circ$, $45^\circ$, and $90^\circ$ with respect to the quantization axis. We stress that the three situations are physically identical---the apparent difference simply results from a transformation between bases. This also means that when diagonalizing the coupling Hamiltonian one will get the same eigenvalues irrespective of the polarization angle. In the dipole and rotating-wave approximations, the coupling is $H=eE(\bm{\epsilon}\cdot{\bf r})/2$ for a resonant microwave field of amplitude $E$ and polarization $\bm{\epsilon}$. Formally, performing a classical rotation $R$ of the polarization vector $\bm{\epsilon}\rightarrow \bm{\epsilon}'$, we have $(\bm{\epsilon}'\cdot{\bf r})=\mathscr{D}^\dagger(R)(\bm{\epsilon}\cdot{\bf r})\mathscr{D}(R)$, where $\mathscr{D}(R)$ is the quantum mechanical rotation operator \cite{Sakurai2020}. Since this constitutes a similarity transformation $(\bm{\epsilon}'\cdot{\bf r})$ has the same eigenvalues as $(\bm{\epsilon}\cdot{\bf r})$.
Specifically, for resonantly coupled $S_{1/2}$ and $ {P}_{1/2}$ levels, the representation of $H$ in the Zeeman basis with a linearly polarized microwave field oriented along the quantization axis (which we take to be the $z$-axis) is
\begin{equation}\label{eq:unrotHamil}
H =\frac{\hbar}{2} \begin{pNiceMatrix}[first-row,last-col]
    \scalebox{0.7}{\ket{s_{1/2},{-\tfrac{1}{2}}}} &  \scalebox{0.7}{\ket{s_{1/2},{\tfrac{1}{2}}}}& \scalebox{0.7}{\ket{p_{1/2},{-\tfrac{1}{2}}}} & \scalebox{0.7}{\ket{p_{1/2},{\tfrac{1}{2}}}} \\
    0 & 0 & -\Omega_0 & 0 & \scalebox{0.7}{\bra{s_{1/2},{-\tfrac{1}{2}}}}\\
    0 & 0 & 0 & \Omega_0 & \scalebox{0.7}{\bra{s_{1/2},{\tfrac{1}{2}}}}\\
    -\Omega_0^* & 0 & 0 & 0 & \scalebox{0.7}{\bra{p_{1/2},{-\tfrac{1}{2}}}} \\
    0 & \Omega_0^* & 0 & 0 & \scalebox{0.7}{\bra{p_{1/2},{\tfrac{1}{2}}}} 
    \end{pNiceMatrix},
\end{equation}
where the ordering of basis states $\ket{\ell_J,m_J}$ used to calculate the matrix elements $\bra{\ell'_{J'},m'_{J'}}\frac{1}{2}eE(\bm{\epsilon}\cdot{\bf r})\ket{\ell_j,m_j}$ are shown at the upper and right margins of the $4\times4$-matrix. For example
\begin{equation}
\begin{split}
H_{13}&=
    \bra{s_{1/2},-\tfrac{1}{2}}{\tfrac{1}{2}eE(\bm{\epsilon}\cdot{\bf r})}\ket{p_{1/2},-\tfrac{1}{2}}\\ & =\tfrac{1}{2}eE\underbrace{\bra{s_{1/2},-\tfrac{1}{2}}{r_0}\ket{{p_{1/2},-\tfrac{1}{2}}}}_{-\begin{pmatrix}
    \tfrac{1}{2} & 1 & \tfrac{1}{2}\\
    \tfrac{1}{2} & 0 & -\tfrac{1}{2}
    \end{pmatrix}\ReducedMat{s_{1/2}}{r}{p_{1/2}}}\\ & =-eE\frac{1}{2\sqrt{6}}\ReducedMat{s_{1/2}}{r}{p_{1/2}}=-\frac{\hbar}{2}\Omega_0,
    \end{split}
\end{equation}
with Rabi frequency $\Omega_0\equiv e E\frac{1}{\sqrt{6}}\ReducedMat{s_{1/2}}{r}{p_{1/2}}/\hbar$. In obtaining this, the Wigner-Eckart theorem has been used to express $\bra{s_{1/2},-\tfrac{1}{2}}{r_0}\ket{{p_{1/2},-\tfrac{1}{2}}}$ as a signed product of of a 3-$j$ symbol and a reduced matrix element (cf. underbraced quantity). Noting that for the reduced matrix element we have $\ReducedMat{s_{1/2}}{r}{p_{1/2}}=\ReducedMat{p_{1/2}}{r}{s_{1/2}}^*$ \cite{Auzinsh2010}, the remaining three non-zero entries of \eref{eq:unrotHamil} follow in a similar fashion. The coupling Hamiltonian $H$ in \eref{eq:unrotHamil} has two eigenvalues $\pm\hbar|\Omega_0|/2$.

Upon rotating the microwave polarization by an angle $\theta$ about the $y$-axis, the Hamiltonian for the coupled system becomes
\begin{equation}\label{eq:rotHamil}
H^\theta =\frac{\hbar}{2} \begin{pmatrix}
    0 & 0 & -\Omega_0\cos\theta & \Omega_0\sin\theta\\
    0 & 0 & \Omega_0\sin\theta & \Omega_0\cos\theta\\
    -\Omega_0^* \cos\theta& \Omega_0^*\sin\theta & 0 & 0 \\
    \Omega_0^*\sin\theta & \Omega_0^*\cos\theta & 0 & 0 
    \end{pmatrix},
\end{equation}
where, for example
\begin{eqnarray}
H_{13}^\theta&=&
    \tfrac{1}{2}eE{\bra{s_{1/2},-\tfrac{1}{2}}\mathscr{D}^\dagger(R)r_0\mathscr{D}(R)\ket{p_{1/2},-\tfrac{1}{2}}}\\ \nonumber
    & =& \tfrac{1}{2}eE\left( \cos\theta/2\bra{s_{1/2},-\tfrac{1}{2}}+\sin\theta/2\bra{s_{1/2},\tfrac{1}{2}}\right)\\ \nonumber
   &&\phantom{xxxx}  r_0(\cos\theta/2\ket{p_{1/2},-\tfrac{1}{2}}+\sin\theta/2\ket{p_{1/2},\tfrac{1}{2}})\\ \nonumber
    & =& \tfrac{1}{2}eE(\cos^2\theta/2\underbrace{\bra{s_{1/2},-\tfrac{1}{2}}r_0\ket{p_{1/2},-\tfrac{1}{2}}}_{-\hbar\Omega_0/eE}\\ \nonumber &&+\sin^2\theta/2\underbrace{\bra{s_{1/2},\tfrac{1}{2}}r_0\ket{p_{1/2},\tfrac{1}{2}}}_{\hbar\Omega_0/eE})=-\frac{\hbar}{2}\Omega_0\cos\theta.
    \end{eqnarray}
\Eref{eq:rotHamil} differs to equation B1 of \cite{Chopinaud2021}, by (at least) a sign in two entries. This explains why Ref.~\cite{Chopinaud2021} reaches the conclusion that the eigenvalues change as the polarization is rotated. In contrast, and as expected from symmetry, $H^\theta$ of \eref{eq:rotHamil} displays the same two eigenvalues as H in \eref{eq:unrotHamil}, namely $\pm\hbar|\Omega_0|/2$.
\section{Experiment}
\subsection{$S_{1/2} \leftrightarrow P_{1/2}$ transition}
\begin{figure}[b!]
    \centering
    \includegraphics[width=\columnwidth]{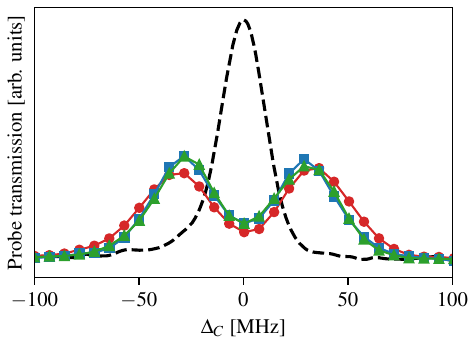}
    \caption{Probe transmission as the coupling laser frequency is scanned. Black dashed line is the bare EIT (no MW field). \textcolor{BrickRed}{\ding{108}}, \textcolor{NavyBlue}{\ding{110}} and \textcolor{ForestGreen}{\ding{115}} are with a MW field oriented at $\theta_{\text{MW}}=\SI{0}{\degree}, \SI{45}{\degree}, \text{and}\ \SI{90}{\degree}$ respectively.}
    \label{fig:thz-on-off}
\end{figure}

\begin{figure}[b]
    \centering
    \includegraphics[width=\columnwidth]{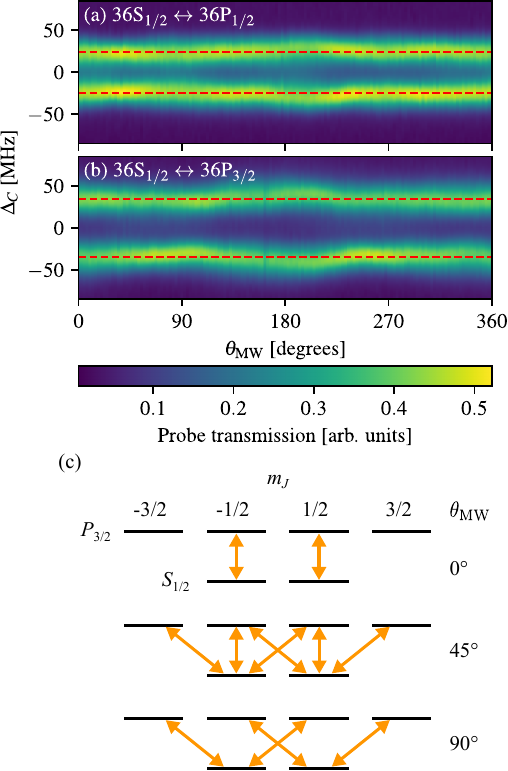}
    \caption{Polarization scans of the (a) $36S_{1/2} \leftrightarrow 36P_{1/2}$ and (b) $36S_{1/2} \leftrightarrow 36P_{3/2}$ transitions as the coupling laser frequency is scanned. Probe transmission is normalized relative to the height of an unperturbed EIT trace; each plot is the average of 10 scans. Two Gaussians were fitted to obtain peak centres; their mean positions are plotted as red dashed lines. (c) $m_J$ magnetic sublevels involved in the $36S_{1/2} \leftrightarrow 36P_{3/2}$ transition, and MW couplings for $\theta_{\text{MW}}=\SI{0}{\degree}, \SI{45}{\degree}, \text{and}\ \SI{90}{\degree}$.}
    \label{fig:pol-scans}
\end{figure}

To experimentally investigate the polarization dependence for an $S_{1/2} \leftrightarrow P_{1/2}$ transition probed using EIT, we perform spectroscopy via the ladder scheme of \fref{fig:pol-spec-diagram}(a) with the experimental configuration shown in \fref{fig:pol-spec-diagram}(c). Here, a \SI{780}{nm} probe laser is locked to the $5S_{1/2}\ (F=2)\leftrightarrow5P_{3/2}\ (F=3)$ transition in \Rb, while a \SI{480}{\nano\metre} coupling laser is tuned to the $5P_{3/2} \leftrightarrow 36S_{1/2}$ transition, establishing an EIT condition for the probe beam by the atomic medium. The coupling laser is scanned, and the EIT transmission is recorded as a function of the coupling laser detuning from the $5P_{3/2}\rightarrow36S_{1/2}$ resonance. The dashed line in \figref{fig:thz-on-off} shows a single transmission peak, centered on zero detuning. The probe and coupling beams have Rabi frequencies of $2\pi \times\SI{6.0}{MHz}$ and $2\pi\times\SI{0.67}{MHz}$, respectively, and $1/e^2$ diameters of \SI{2.0}{mm} and \SI{1.9}{mm}. We modulate the amplitude of the coupling beam at \SI{10}{\kilo\hertz} using an optical chopper, enabling the use of lock-in detection. The probe and coupling beams counter-propagate along the $z$-axis to suppress effects of Doppler broadening and are circularly polarized. \Fref{fig:thz-on-off} shows the effect on the EIT spectrum of applying a linearly polarized MW field tuned to the $36S_{1/2} \leftrightarrow 36P_{1/2}$ transition at \SI{86.355}{GHz} and propagating along the $y$-axis \footnote{Our experimental implementation uses a microwave field produced by a commercially available terahertz source (Toptica TeraBeam 1550), with the emitter mounted in a motorized rotation stage positioned \SI{13}{cm} above the cell [cf. \fref{fig:pol-spec-diagram}(c)].}. Near-identical AT splitting profiles are found for the polarization angles $\theta_{\text{MW}}=0^\circ,45^\circ,90^\circ$---corresponding to the situations of \fref{fig:pol-spec-diagram}(b). This invariance of peak heights and splittings extends to the entire angular range
of $\theta_{\text{MW}}$.
\Fref{fig:pol-scans}(a) maps the AT splitting as \thetaMW\ is rotated. EIT spectra are recorded every \SI{2.5}{\degree}, all showing two peaks. The peak positions remain constant within $3$~MHz from their angular averages, overlaid on \fref{fig:pol-scans}(a) as two red dashed lines split by \SI{49}{MHz}. For a resonant MW field, this frequency splitting is related to the MW Rabi frequency $\Omega_{\text{MW}}$ by~\cite{Sedlacek2012}
\begin{equation}\label{eq:power}
    \Delta f = \frac{\Omega_{\text{MW}}}{2\pi},\quad \Omega_{\text{MW}}=\frac{\mu_{ij}E}{\hbar},
\end{equation}
where $\mu_{ij}$ is the transition dipole moment between states \ket{i} and \ket{j}, and $E$ is the electric field amplitude. We attribute the small variation in the observed AT splitting to the MW field distribution not being cylindrically symmetric about the axis of rotation~\cite{Nellen2021, Smith2021}.
\subsection{$S_{1/2} \leftrightarrow P_{3/2}$ transition}
So far we have investigated a linear MW field dressing a 
$J=1/2\leftrightarrow J'=1/2$ system and established that this gives rise to two eigenenergies for the coupling Hamitonian and in turn two peaks in the EIT spectrum. Moreover, the positions and heights of these peaks remain constant as the MW polarization angle is scanned.
In general, the number of unique eigenenergies for a coupled system $\ket{J} \leftrightarrow \ket{{J'}}$ (subject to dipole transition selection rules) can be shown to be (see Appendix \ref{sec:neig})
\begin{equation}
N_{\rm eig}=
\begin{cases}
    J+J'+1, & 2J \text{ odd}\\
    J+J'+2, & 2J \text{ even, } J\ne J'\\
    J+J'+1, & 2J \text{ even, } J=J'    
\end{cases}.\label{eq:neigen}
\end{equation}
We stress that the $N_{\rm eig}$ eigenenergies are polarization insensitive.
For $J=J'$, the number of (stationary) peaks in the EIT spectrum equals $N_{\rm eig}$, as for the $S_{1/2}\leftrightarrow P_{1/2}$ case considered above. For $J\neq J'$, the number of peaks will depend on which of the two the EIT coupling laser connects to (see Appendix \ref{sec:npeaks}). If the last optical rung on the EIT ladder corresponds to the larger of $J$ and $J'$ the peak number equals $N_{\rm eig}$. However, if the coupling laser light connects to the level with the lower of the two, the number of peaks will be $N_{\rm eig}-1$. Hence if for the topmost level of the ladder \fref{fig:pol-spec-diagram}(a) we make use of $36P_{3/2}$ rather than $36P_{1/2}$, we would still expect only two peaks in the EIT spectrum despite the $S_{1/2}\leftrightarrow P_{3/2}$ coupling matrix having three eigenvalues. Indeed, \fref{fig:pol-scans}(b) presents results from experimental measurements with the microwave field tuned in resonance with the $36S_{1/2}\leftrightarrow 36P_{3/2}$ transition at \SI{88.697}{GHz}, revealing two peaks split by \SI{70}{MHz}. This splitting is $\sim 40~\%$ larger than for the $36S_{1/2}\leftrightarrow 36P_{1/2}$ transition and since the difference in transition dipole moments for the two is only $\sim 2\%$ this must be the direct result of a change in MW power at the atoms' location [cf.~\eref{eq:power}]. We ascribe the frequency dependence of the emitted power to the THz source. As such, the polarization-insensitivity of the $S\leftrightarrow P$ transitions of Figs.~\ref{fig:pol-scans}(a) and \ref{fig:pol-scans}(b) makes this experimental scheme a promising candidate for measuring emitter properties. 
\section{Discussion and Conclusion}
\Fref{fig:pol-scans}(c) shows how the $36S_{1/2}$ and $36{P}_{3/2}$ magnetic sublevels are coupled for linearly polarized MW at angles $0^\circ$, $45^\circ$, and $90^\circ$ with respect to the quantization axis. For the $0^\circ$ case it is straightforward to see that the system will have three eigenenergies, as it is represented by two independent, identical, coupled two-level systems in addition to two uncoupled energy-degenerate states, coined spectator states \cite{Shore2013}. With the coupling laser connecting to the $S_{1/2}$ level it is clear that the spectator states will have no influence on the EIT spectrum. For the representations of the $45^\circ$ and $90^\circ$ cases, which both contain linkages between $>2$ states, the number of eigenenergies is not obvious from inspection of the diagrams in \fref{fig:pol-scans}(c). The Morris-Shore transformation \cite{Shore2013} provides the Hilbert space coordinate transforms that will turn the linked systems into the unlinked one. While this method lends itself to more general polarization states \cite{Bevilacqua2022}, for the purely linearly polarized radiation considered in the present work this is simply achieved by picking a quantization axis coaxially with the polarization.

As stated above, the eigenenergies found when coupling two Ryd\-berg levels with linearly polarized MW are universally insensitive to the polarization angle. In turn this means that the peak positions for an EIT signal probing a level with splittings remain constant. The prominence of each (stationary) peak will, however, vary in general.  The invariance of the EIT signals with angle in \fref{fig:pol-scans}(a,b) are special cases of the coupling light connecting to a $J=1/2$ level and the associated symmetry. Stationary eigenenergies are a necessary but not sufficient condition for the invariance, and in general the EIT signal can carry an imprint of the polarization angle, as demonstrated in \cite{Sedlacek2013} for a $S_{1/2}\leftrightarrow P_{3/2}\leftrightarrow D_{5/2}\leftrightarrow P_{3/2}$ ladder. In a future publication we will contrast a $S_{1/2} P_{3/2}D_{5/2}P_{3/2}$-ladder to a $S_{1/2} P_{3/2}D_{3/2}P_{1/2}$-ladder. These two scenarios display distinctly different angular EIT signatures and may form the basis of a differential polarimetry scheme for MW and THz radiation. Meanwhile, $S\leftrightarrow P$ Rydberg transitions, that were the focus of the present paper, can, as we have demonstrated, not be used for polarimetry. Rather, such transitions should be sought in applications where insensitivity to polarization effects are desired. 
\begin{acknowledgments}
This work was supported by the Marsden Fund of New Zealand (Contracts Nos. UOO1923 and UOO1729) and by MBIE (Contract No. UOOX1915). NK acknowledges the hospitality of Aarhus University during the write-up of the manuscript. We thank Jevon Longdell for comments on our manuscript.
\end{acknowledgments}
\appendix
\begin{figure}[tb!]
	\centering
	\includegraphics[width=\columnwidth]{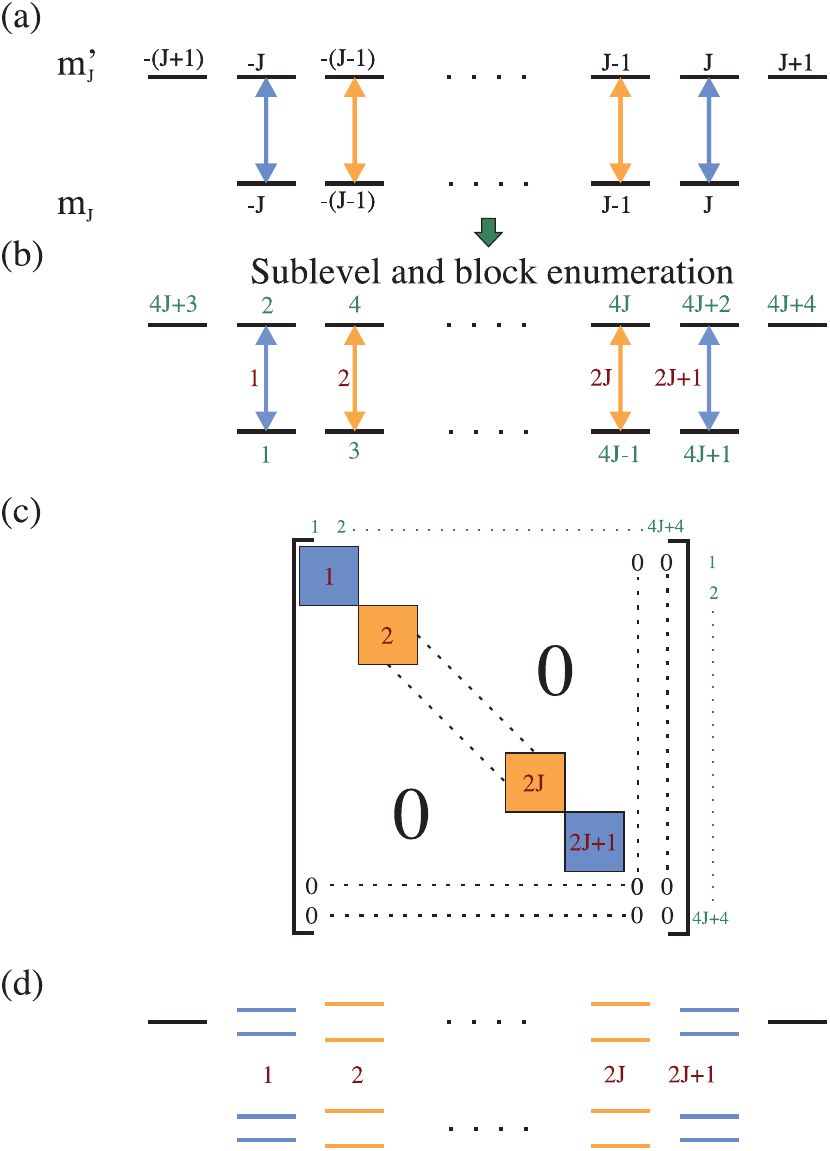}
	\caption{(a) Magnetic sublevels $m_J$ and $m'_{J}$ for a $J\leftrightarrow J'$-transition with $J'=J+1$, driven with a linearly polarized field; the quantization axis has been chosen along the polarization. By enumerating the sublevels (green numbers) and transitions (red numbers) as shown in (b), the coupling Hamiltonian will be represented by the block-diagonal matrix in (c). Each allowed transition $1,2... 2J+1$ corresponds to a $2\times 2$ block on the diagonal of this coupling  matrix. (d) Dressed state representation of the coupled $J\leftrightarrow J'$ system.}
	\label{fig:appfig}
\end{figure}
\section{Number of coupling Hamiltonian eigenvalues}\label{sec:neig}
Here we underpin \eref{eq:neigen} which states the number of unique eigenenergies $N_{\rm eig}$ for a system $\ket{J} \leftrightarrow \ket{{J'}}$ coupled by a linearly polarized field.

The selection rules for dipole transitions implies $\Delta J=0,\pm1$ \cite{Foot2005}, and we  first consider the case $\Delta J=1$, so that $J'=J+1$.  \Fref{fig:appfig}a shows a schematic level diagram indicating the\nocite{3j}\nocite{Edmonds1957} allowed $\Delta m_J=0$ transitions \cite{3j}. The system has a total of $4J+4$ states which we will enumerate as indicated in \fref{fig:appfig}b. Prescribing the ordering of states using this particular enumeration, the coupling matrix becomes block diagonal (see \fref{fig:appfig}c) with $2J+1$ non-zero Hermitian $2\times 2$ blocks, each corresponding to an allowed $\Delta m_J=0$ transition designated by red numbering in \fref{fig:appfig}b. Furthermore, the coupling matrix contains one $2\times 2$ null-matrix block arising from the stretched spectator states of the $J'$-level. Each of the non-zero blocks is itself off-diagonal and has two non-zero eigenvalues \footnote{ The two eigenvalue have the same the magnitude, differing by a sign, as their sum must equal the zero trace of the off-diagonal $2\times2$ block.}. We note the symmetry of the system \cite{3j}: transition 1 of \fref{fig:appfig}b has the same strength as transition $2J+1$, transition 2 has the same strength as transition $2J$, etc. This ensures that the corresponding pairs of coupling matrix blocks will share eigenvalues. Incorporating this degeneracy, the non-zero blocks amount to $2J+1=J+J'$ unique eigenvalues if the number of blocks is even ($2J$ odd) and $2J+2=J+J'+1$ unique eigenvalues if the number of blocks is odd ($2J$ even). When including the zero eigenvalue of the null-block the total number of eigenvalues becomes $J+J'+1$ and $J+J'+2$ [cf. \eref{eq:neigen}, first and second case], respectively. These results remain valid if $\Delta J=-1$.

For $\Delta J=0$\nocite{3j2} \cite{3j2}, there are $2J+1= J+J'+1$ dipole-allowed transitions if $2J+1$ is even ($2J$ odd) and hence $J+J'+1$ unique non-zero eigenvalues, and there is no zero eigenvalue from a null-block contributing to $N_{\rm eig}$ due the absence of stretched spectator states. Finally, for $\Delta J=0$ and $2J+1$ odd ($2J$ even), there are only $2J=J+J'$ allowed transitions, because $m_J=0\leftrightarrow m_{J}'=0$ is dipole forbidden when $J=J'$ \cite{Foot2005}. Hence the coupling matrix will possess a zero eigenvalue in addition to the $J+J'$ non-zero eigenvalues so that $N_{\rm eig}=J+J'+1$ [cf. \eref{eq:neigen}, third case].
\section{Number of EIT spectrum peaks }\label{sec:npeaks}
\Fref{fig:appfig}(d) presents the coupled system of Figs.~\ref{fig:appfig}(a) and \ref{fig:appfig}(b) in a dressed state picture \cite{CohenTannoudji1996}. Centered on the lower $J$-level is a series of Autler-Townes doublets, each split by an energy equal to the difference between the two eigenvalues of the corresponding coupling block.
Probing this level as the upper rung in an EIT ladder scheme would give rise to spectrum with a number of peaks equal to $N_{\rm eig}-1$. The corresponding spectrum for the upper $J'$-level would additionally include a degenerate peak from the two uncoupled, unsplit spectator states (black lines) and hence contains $N_{\rm eig}$ peaks.

%

\end{document}